\documentstyle[aps,prl,multicol,epsf]{revtex}
\def\beq{\begin{equation}} \def\eeq{\end{equation}}
\def\bea{\begin{eqnarray}} \def\eea{\end{eqnarray}}

\def\beann{\begin{eqnarray*}} \def\eeann{\end{eqnarray*}}

 \let\z=\zeta \let\h=\eta 
\let\eps=\epsilon
  \let\la=\lambda 
  \let\p=\pi  \let\s=\sigma
 \let\ps=\psi
\let\ph=\varphi

\let\qd=\quad  

\def\tst#1{{\textstyle #1}}

\def\0{\over } \def\1{\vec }     \def\2{{1\over2}} \def\4{{1\over4}}
\def\5{\bar }  \def\6{\partial } \def\7#1{{#1}\llap{/}}

\def\<{\langle } \def\>{\rangle }

\let\auf=\uparrow \let\ab=\downarrow

\let\nodoti\i
\def\i{{\rm i}}

\def\sign{\mbox{sign}}  
\def\mod{\mbox{\,mod\,}}

\input{epsf}
\begin{document}

\title{Charge and spin separation in the $1D$ Hubbard model}
\author{Vladimir E. Korepin$^\dagger$ and Itzhak 
Roditi$^{\dagger\ddagger}$}
\address{${}^{\dagger}$C. N. Yang Institute for Theoretical Physics,\\
State University of New York at Stony Brook, NY 11794-3840, U.S.A.\\
${}^{\ddagger}$Centro Brasileiro de Pesquisas F\'{\nodoti}sicas, \\
Rua Dr. Xavier Sigaud 150, 22290-180, Rio de Janeiro, RJ - Brazil.}

\date{\today}

\maketitle

\begin{abstract}
We consider the repulsive Hubbard model in one dimension and show
the different mechanisms present in the charge and spin separation
phenomena for an electron, at half filling and bellow half
filling. We also comment recent experimental results.
\end{abstract}

\draft
\pacs{PACS numbers: 71.10.Fd 71.27.+a 71.30.+h 74.20.-z}

\begin{multicols}{2}

Electrons possess charge and spin, as strong as this unified
picture may be imprinted in our minds these two quantities have
been discovered separately. It is possible to trace back the
discovery of the electron charge from the first experiments of
J.J. Thomson to the measurements of R. Millikan that demonstrated
the existence of a fundamental electric charge. Only sixteen years
later G. E. Uhlenbeck and S. A. Goudsmith discovered that the
electron not only has a charge, but also carries spin. The
electron was the first elementary particle discovered. It is hard
to recall any discovery, since then, with such an impact, not only
on physics but on science and technology as well.

Since the first proposals \cite{Gutz,hub,kan}, of a model of a local
repulsive
interaction as an appropriate system
for modelling magnetism and metal-insulator transitions the so called
Hubbard model became the prototype for the study of
strongly correlated electrons.

Strongly correlated electronic systems are under intense
investigation, due to their possible role in high-$T_c$
superconductivity. In particular, the two dimensional Hubbard
model is in the very center of those investigations, but one
cannot find an exact solution and many results must rely on
numerical calculations. On the other hand the one dimensional
Hubbard model is an integrable model and it is believed that some
of its most important features carry over higher dimensions
\cite{ander1}. Also the most outstanding physical phenomenon
happens in the $1D$ Hubbard model. Charge separates from spin.

The Hamiltonian of the one-dimensional Hubbard model on a periodic
$L$-site lattice may be written as
\bea
     H = - \sum_{j=1}^L \sum_{\s = \auf, \ab}
       (c_{j, \s} + c_{j+1, \s} + c_{j+1, \s}^+ c_{j, \s})
\nonumber \\
            + U \sum_{j=1}^L
       (n_{j \auf} - \tst{\2})(n_{j \ab} - \tst{\2}) .
\label{ham}
\eea
$c_{j, \s}^+$ and $c_{j, \s}$ are canonical creation and annihilation
Fermi operators,
  $n_{j,\s} = c_{j, \s}^+
c_{j,\s}$ is the particle number operator for electrons of spin $\sigma$ in the
Wannier state centered around the site $j$ and $U$ is the coupling
constant. Periodicity is guaranteed by
setting $c_{L+1, \s} = c_{1, \s}$.
The Hubbard Hamiltonian
conserves the number of electrons $N$ and the number of down spins $M$,
Lieb and Wu \cite{LiWu} solved it, for fixed $N$ and $M$
by means of the nested Bethe
ansatz \cite{Yang67}. They calculated the ground state energy at half
filling, $N/L=1$. A review of the main results in the Hubbard model can be
found in \cite{reprK}. Some of the important symmetries of this
Hamiltonian are its invariance
under particle-hole transformations, under reversal of spins
\cite{LiWu}, allowing one to set $2M \le N \le L$.
In the following we shall denote the positions and spins of the electrons
by $x_j$ and $\s_j$, respectively.
The Bethe ansatz eigenfunctions of the Hubbard Hamiltonian (\ref{ham}) in
a sector $Q$ are given as
\bea
&\ps (x_1, \dots, x_N; \s_1, \dots, \s_N)&= \nonumber  \\
&\sum_{P \in S_N}
 \sign(PQ) \, \ph_P (\s_{Q1}, \dots, \s_{QN})&
\exp \left( \i \sum_{j=1}^N k_{Pj} x_{Qj} \right).
 \label{wwf}
\eea
$Q$ is a permutation related to one of the $N!$ possible orderings of the
coordinates of $N$ electrons.
The $P$-summation extends over all permutations of the numbers
$1, \dots, N$, these permutations form the symmetric group $S_N$ and the
function $\sign(Q)$ is the sign function on the symmetric group,
which is $- 1$ for odd permutations and $+ 1$ for even permutations.
The spin dependent amplitudes
$\ph_P (\s_{Q1}, \dots, \s_{QN})$
are given in Woynarovich \cite{Woynarovich82a} and have
form the Bethe ansatz wave functions
of an inhomogeneous XXX spin chain. Two sets of
quantum numbers characterize the wave functions (\ref{wwf}) the charge
momenta $\{k_j\}$ and the spin rapidities $\{\la_l\}$. They satisfy the
Lieb-Wu equations
\bea \label{bak}
     e^{\i k_j L} & = & \prod_{l=1}^M \frac{\la_l - \sin k_j - \i U/4}
                                      {\la_l - \sin k_j + \i U/4} \qd,
                              \nonumber \\
                j & = & 1, \dots, N \qd, \\
                      \label{bas}
     \prod_{j=1}^N \frac{\la_l - \sin k_j - \i U/4}
                        {\la_l - \sin k_j + \i U/4} & = &
     \prod_{m=1 \atop m \ne l}^M \frac{\la_l - \la_m - \i U/2}
                        {\la_l - \la_m + \i U/2} \qd,
            \nonumber \\
           l & = &  1, \dots, M \qd.
\eea

The wave functions (\ref{wwf}) are joint eigenfunctions of the
Hubbard Hamiltonian (\ref{ham}) and the momentum operator
with eigenvalues
\bea
     E &=& - 2 \sum_{j=1}^N \cos k_j + \frac{U}{4}(L - 2N) \qd, \qd
\nonumber \\     P &=& \left( \sum_{j=1}^N k_j \right) \mod \, 2\p \qd.
\label{enmom}
\eea

Another crucial
property, for the study of the spectrum, is the invariance of
(\ref{ham}) under the $so(4)$ symmetry
\cite{LiHei}.
The Hubbard Hamiltonian (\ref{ham}) is invariant under rotations in
spin space. The corresponding su(2) Lie algebra is generated by the
operators
\bea
\z &=& \sum_{j=1}^L c_{j \auf}^+ c_{j \ab} ~,~
\z^\dagger = \sum_{j=1}^L c_{j \ab}^+ c_{j \auf}  \nonumber \\
\z^z &=& \tst{\2} \sum_{j=1}^L (n_{j \ab} - n_{j \auf}),
\label{rot}
\eea
where $[\z,\z^\dagger] = - 2 \z^z$~,~$[\z,\z^z] = \z$ and
$[\z^\dagger,\z^z] = - \z^\dagger$.

When the  length $L$ of the lattice is even there is another
representation of su(2), which commutes with the Hubbard Hamiltonian
\cite{LiHei} and generates the so-called $\h$-pairing symmetry,

\bea
\h &=& \sum_{j=1}^L (- 1)^j c_{j \auf} c_{j \ab} ~,~
\h^\dagger = \sum_{j=1}^L (- 1)^j c_{j \ab}^+ c_{j \auf}^+ \nonumber \\
\h^z &=& \tst{\2} \sum_{j=1}^L (n_{j \ab} + n_{j \auf}) ,
 \label{eta}
\eea
where $[\h,\h^\dagger] = - 2 \h^z$ ~,~
        $[\h,\h^z] = \h$ and
        $[\h^\dagger,\h^z] = - \h^\dagger$ .

The role of the $\h$-pairing symmetry is to connect sectors of the
Hilbert space with different numbers of electrons.
The generators of both the rotations in spin space and
$\h$-pairing symmetry commute with one-another and combine
into a representation of $su(2) \otimes su(2)/ Z_2 = so(4)$. Making use
of this algebraic structure it was shown that the set of
Bethe-ansatz eigenstates can be extended leading to a complete set of
eigenstates \cite{EKS92}.

As a result from the above $so(4)$ symmetry it is possible to classify the
excitation spectrum over the half-filled ground state, with respect to
its representations. The charge and spin degrees of freedom
separate. The scattering states of only four elementary excitations
provide, in the repulsive case, the whole spectrum of excitations. Two of
these elementary excitations transform as a triplet
under the spin transformations and as a singlet under the
$\h$-pairing. This means that they carry spin but no charge. The resulting
representation can be denoted by $(\frac{1}{2},0)$ and one calls
them spinons. The $(0,\frac{1}{2})$ representation of $so(4)$ is
associated to the other
two elementary excitations, the holon and the antiholon, which carry
charge but no spin. Bellow half filling there are infinitely many
excitations, although only few of them are essential for low energy
physics.

At half filled band the energy and momentum of the spinons are expressed
in terms of a spectral parameter $\Lambda$, which plays a role similar to
the rapidity for relativistic particles. This excitation is gapless and
the dispersion curves for
the energy  $\eps_s(\Lambda)$ and momentum $p_s(\Lambda)$ are

\bea
\eps_s(\Lambda) &=&
2\int_0^\infty\frac{d\omega}{\omega}\frac{J_1(\omega)\
\cos(\omega\Lambda)}{\cosh(\omega U/4)}\ ,\\ \nonumber
p_s(\Lambda) &=& -\pi/2 -  \int_{0}^\infty {d\omega\over
\omega}
{J_0(\omega) \cos(\omega\Lambda)\over {\cosh}(\omega U/4)},
\eea
where $0~<~p_s(\Lambda)~<~\pi$ and $J_{0,1}$ are Bessel functions.

The holon and antiholon momenta differ by $\pi$ as $\h^\dagger$ does not
commute with the momentum operator changing it by $\pi$. Their dispersion
curves \cite{Woynarovich82a} are parametrized by $k^h$ the bare momentum
of the holon and the spectral parameter is now $sin(k^h)$. For the holon,
which has a gap,
we have,

\bea
\eps_h(\Lambda) = \frac{U}{2} + 2 cos(k^h)~~~~~~~~~~~~~~~~~~~~~~~~ \quad
\\
\nonumber
+ 2\int_0^\infty\frac{d\omega}{\omega}\frac{J_1(\omega)\
\cos[(\omega\sin(k^h)] e^{-\omega U/4}}{\cosh(\omega U/4)}\ ,\quad \\
\nonumber
p_h(\Lambda) = \pi/2 - k^h  \int_{0}^\infty {d\omega\over
\omega}
{J_0(\omega) \cos(\omega\Lambda)\over {\cosh}(\omega U/4)}. \quad
\eea

Applying the operator $\h^\dagger$ the energy and momentum of the
antiholon are found to be $\epsilon_{ah}(k)=\epsilon_h(k), p_{ah}(k)= -\pi
+ p_h(k)$.

Together the Lieb-Wu equations and the $so(4)$ symmetry give a complete
description of all the excitations. The usual approach is to determine
the Bethe ansatz excitation taking the zero-temperature limit of the
thermodynamic Bethe ansatz, making use of the string hypothesis. The Bethe
states are then parametrized by sets of spectral parameters
$\{k_j\},\{\Lambda_a\},\{\Lambda_a\prime\}$, which are subject to a set of
coupled algebraic equations \cite{taka}. The scattering states of these
excitations provide the two body scattering matrix. In \cite{kor80} it was
explained how to deal with the non trivial structure of a ground state
related to  many-body wave functions of Bethe ansatz solvable models in
order to obtain the scattering matrix. For the Hubbard model the complete
scattering matrix is a block diagonal $16\times16$ matrix, where each
$4\times4$ block describes the different scattering states such as the,
spinon-spinon, spinon-holon, holon-spinon and holon-holon.

Let us now examine the charge and spin separation at the half filled band
in more detail. We first remark the differences in the excitation spectrum
in the free, $U=0$, and interacting, $U>0$, cases. When $U=0$ the
electron is the only excitation, it carries charge and spin. In
the interacting case the electron disappears from the spectrum and
we have instead the spinons, which carry spin, and the holons,
which carry charge together with the antiholons. When $U>0$ the
low energy dispersion curve for the spinons is linear, indicating
gapless excitations, but the holon and antiholon dispersion curve
is quadratic an the excitation has a gap. There are no other
excitations at any energy. After separation occurs the spinon and
holon still interact. The scattering state of a spinon holon can be
constructed by choosing $N=L-1$ and
$M=\frac{L}{2} - 1$, which means that one hole is introduced in the
$k_j$'s and
$\Lambda_\alpha^1$'s distributions. The spin and $\h$-pairing of
this excitations are given by $S=\frac{1}{2}=\h$ and
$S^z=-\frac{1}{2}=\h^z$. The scattering states of two spinons and
two holons can also be constructed \cite{kor2}. Knowledge of those
states allows the calculation of the scattering matrix $S$ of a
spinon with momentum $p$ on the standing holon as,
\begin{equation}\label{smat}
\frac{p+i 2 J_0(i\frac{\pi}{2 U})}{p-i 2 J_0(i\frac{\pi}{2 U})}.
\end{equation}
In the above expression the term, $2 J_0(i\frac{\pi}{2 U})$, is the
interaction constant between the holon and spinon and $J_0$ is a Bessel
function.

Bellow half filling the situation is quite diverse, one can calculate the
spin and charge velocities and realize that they are different. Meaning
that if we drop an electron on the chain the spin will fly away from
the charge. From another point of view one can consider the spectrum of
low lying excitations and verify that it is possible to classify them into two
categories, $i)$ possessing charge but no spin and gapped or gapless,
$ii)$ possessing spin but no charge and gapless. The electron state
being gapped.

Yet another possibility to approach the charge and spin separation
property is to consider correlation functions at finite temperature $T$:

\begin{equation} \label{tcorr}
g(x,t,T)=\frac{
tr\{
e^{-\frac{H}{T}}c^{\dagger}_{\uparrow}(x,t)c_{\uparrow}(0,0)\}
                                                   }
{tr\{e^{-\frac{H}{T}}\}}
\end{equation}

Long distance asymptotics of these correlation functions is
governed by the conformal dimensions $\triangle^{\pm}$. In the
case of the Hubbard model the conformal dimensions
$\triangle^{\pm}_{c,s}$ associated to the charge and spin
distributions have been calculated in \cite{frkor}. It is then
possible to realize that, asymptotically, the correlation function
factorizes into two factors a charge and a spin one, which we find
to be given by,
\begin{equation}\label{factor}
\prod_{\pm} exp\{\frac{-2\pi T}{v_c}\triangle^{\pm}_c |x \mp v_c t|\}
exp\{\frac{-2\pi T}{v_s}\triangle^{\pm}_s |x \mp v_s t|\},
\end{equation}
where $v_{c}$ and $v_{s}$ are the Fermi velocities for the charge and spin
excitations. This, in turn, implies the factorization of the creation
operators of an electron $c^{\dagger}_{j,\sigma}$ into spin and charge
factors.

We have thus discussed the phenomenon of charge
and spin separation in the $1D$Hubbard
model. With the present exact results we expect to provide a basis
for the usual intuition developed in the studies of strongly correlated
electron systems (see for instance
\cite{ander2}).

It is worth noticing that numerically charge and spin separation was shown
in some extensions of the Hubbard model \cite{balseiro}. Also in the
framework of the $t-J$ model charge and spin separation occurs
\cite{bares}. In the two dimensional realm a theoretical proposal, 
exploring a discrete gauge symmetry, to investigate
charge and spin separation has been put forward by Senthil and Fisher 
\cite{fisher}.

More recently, several groups started experimental verification of
charge and spin separation using the technique of angle-resolved
photoemission spectroscopy (ARPES) in quasi one-dimensional
materials. ARPES experiments can provide detailed
dispersion curves through the analysis of the energy and emission
angle of electrons hit by monochromatic photons. Using this
technique one group \cite{kim} studied the $1D$ charge-transfer
insulator ${Sr}{Cu}{O_2}$, another experimental group studied the
$1D$ Mott-Hubbard type insulator ${Na}{V_2}{O_5}$
\cite{kobayashi}. The results coming from both experiments require
a theoretical framework where the charge and spin degrees of
freedom separate.

{\bf Acknowledgements:} V.~K. is grateful to NSF support under grant
PHY-9605226. I.~R. would like to thank CNPq (Conselho Nacional
de Desenvolvimento Cient\'{\nodoti}fico e Tecnol\'ogico) for financial
support.

%


\end{multicols}

\end{document}